\documentclass[aps,prl,amsmath,twocolumn]{revtex4}
\usepackage{bm}
\usepackage{graphicx}

\def \d{\partial}
\def \bv{{\bf v}}

\def \bomega{\boldsymbol{\omega}}

\begin{document}

\title{Lagrangian  and Eulerian velocity structure functions in hydrodynamic
turbulence}

\author{ K.P. Zybin
\footnote{Electronic address: zybin@lpi.ru}, V.A. Sirota}

\affiliation{ 119991 P.N.Lebedev Physical Institute, Russian Academy of Sciences,
Moscow, Russia}

\begin{abstract}
The Lagrangian  and Eulerian transversal velocity structure
functions of fully developed fluid turbulence are found basing on
the Navier-Stokes equation. The structure functions are shown to
obey the scaling relations $S_n^L(\tau)\propto\tau^{\xi_n}$ and
$S_n^E(l)\propto l^{\zeta_n}$ inside the inertial range. The
scaling exponents $\zeta_n$ and $\xi_n$ are calculated
analytically without using dimensional considerations. This result
is a significant step toward understanding the nature of
hydrodynamic turbulence. The obtained values are in a very good
agreement with recent numerical and experimental data.

\vspace{0.4cm} \noindent {\small PACS numbers: 47.10.ad, 47.27.Jv}
\end{abstract}

\maketitle

Understanding of statistical properties of a fully developed
turbulence from the Lagrangian and Eulerian points of view has
been a challenging theoretical and experimental problem for many
years \cite{Frisch,Lvov,Sreenivasan}. Even a bridge relation
between Lagrangian and Eulerian structure function exponents is a
theoretical problem. Also, there has been no theory based on the
solution of the Navier-Stokes equation up to now. Recent progress
in numerical calculations \cite{Bif}-\cite{Benzi} and experiments
\cite{Toschi} encourages to develop a statistical theory. First
successful step in this direction was undertaken in
 \cite{PRL} -\cite{JETP2}. The aim of this paper is to
elaborate the theory of Lagrangian and Eulerian turbulence based on the
Navier-Stokes equation. We consider  fully developed turbulence in incompressible
fluid in non-viscous limit  and advance substantially the  vortex filament model proposed
previously in \cite{PRL}.
On the basis of this model we find scaling exponents of velocity structure functions
and derive a bridge relation between the Lagrangian and Eulerian scaling exponents.

The statistics of turbulence deals with different kinds of
structure functions. In what follows we are interested in
Lagrangian velocity structure functions
\begin{equation}\label{SnLagr}
S_n^L(\tau)=\left< |{\bf v}(t+\tau) - {\bf v}(t)|^n\right>
\end{equation}
and Eulerian transversal structure functions
\begin{equation}\label{SnEuler}
S_n^{\perp}(l)=\left< \left| \left({\bf v}({\bf r}+{\bf l}) - {\bf
v}({\bf r})\right)\times \frac{\bf l}{l}\right| ^n\right>
\end{equation}
In the first ones, the velocity difference is taken along the
trajectory of one particle and the average is taken over the
ensemble of particles. In the second, the velocity difference is
taken in different points at the same time, and averaged over all
pairs of points.  Since the turbulence is assumed to be
stationary, both results must not depend on time. We restrict our
consideration by the values $\tau, l$ that belong to the inertial
range of scales. The usual assumption is that inside the inertial
range the structure functions do not depend on geometry of a flow,
and have a finite limit as the Reynolds number goes to infinity.

In the previous  papers we proposed a theory of vortices formation
built on the Navier-Stokes equation \cite{JETP1,JETP2}. It allowed
us to calculate the Lagrangian velocity structure functions of
different orders \cite{PRL,JETP2}. In this paper we develop and
generalize the theory, analyzing a boundary of a vortex filament
and introducing a notion of a filament's age $t_*$. This allows to
calculate the transversal Eulerian structure functions and to find
a relation between these two types of structure functions.

The basic idea of our theory is that vortex filaments, i.e. the
elongated regions of  high vorticity, make the main contribution
to structure functions.$^1$
\footnotetext[1]{Besides, this
corresponds to the Kolmogorov's law. Actually, from $\delta v
\propto l^{1/3}$ it follows $\frac{\omega}{\omega_0} \simeq \left(
\frac{\delta l}{L_0} \right)^{-2/3}$. In the inertial range
$\delta l \ll L_0$, hence $\omega \gg \omega_0$.}
Thus, not decay
of eddies but their stretching is the main process responsible for
the formation of structure functions. We will see that studying
this stretching allows to obtain 'intermittent' scaling exponents
practically coinciding with experimental and numerical results.

Below, we recall briefly the basic points of our theory, then we
present without derivation the results of the analysis of the
basic equation and generalize our previous results for any (not
only $\delta$-correlated) random process. Then we apply the
results to find the structure functions.

Following \cite{JETP1}, we decompose the spatial derivative of
velocity into the sum of symmetric and asymmetric parts:
$$\d_i v_j=\frac 12 \epsilon_{ijk} \omega_k + b_{ij} \ , b_{ij}= b_{ji}$$
Here $\omega_k$ is the vorticity, and $b_{ij}$ is a traceless
(because of incompressibility) symmetric tensor. Differentiating
the Euler equation, we then get the equations for all the
components of $\omega_i$ and $b_{ij}$. Changing to the
quasi-Lagrangian frame moving along the trajectory of a fluid
particle, we find the ordinary differential equations describing
the evolution of $\omega_i$ and $b_{ij}$ along the trajectory:
\begin{equation}  \label{quasi}
\begin{array}{l}
\dot{b}_{ij} + \frac 14 (\omega_i \omega_j - \omega^2
\delta_{ij}) +b_{ik} b_{kj} + \rho_{ij} =0 \ ,
\\
\displaystyle \dot \omega_n  = b_{nk} \omega_k
 \end{array}
  \end{equation}
Here  $\rho_{ij}=\nabla_i \nabla_j p$, $p$ is the pressure.  The
time derivative of the second equation gives
\begin{equation}
\ddot{\omega}_n=-\rho_{nk}(t) \omega_k  \label{omega}
\end{equation}
along the particle trajectory.
This equation is  a direct consequence
of the Navier-Stokes equation under the limit $\nu\to 0$ without
any additional assumptions.

Consider a vortex filament where vorticity is high and the
characteristic radius $R$ of the filament is much smaller than the
scale of the largest vortices $L$ comparable to the size of the
flow:
\begin{equation} \label{biggradient}
R \sim \left( \nabla \omega/ \omega \right)^{-1} \ll L
\end{equation}
In general, $\rho_{nk}$ is some complicated integral function of
$\omega$. We have shown that inside a vortex filament $\rho_{nk}$
is independent on the local value of $\omega$, and the equation
(\ref{omega}) becomes {\it linear}.

 We now introduce randomness into the equation.
Note that in (\ref{quasi}) we have 3+6=9 differential equations
for 3+6+6=15 values $\omega_i$, $b_{ij}$ and $\rho_{ij}$. The
condition of incompressibility gives one additional algebraic
equation $b_{ii}=0$. Hence, both in (\ref{quasi}) and
(\ref{omega}) we have five free functions of time. So, we treat 
$\rho_{nk}$ as random functions satisfying
$$ \langle \rho_{ij}(t) \rho_{nk}(t') \rangle = D_{ijnk}(t-t') $$
Since the flow is assumed to be locally isotropic, 
$$ D_{ijnk} (\delta t)=
(\delta_{in}\delta_{jk}+\delta_{ik}\delta_{jn})D(\delta t) \ .$$
Normalizing time, we choose $\int D(\tau) d \tau =1$.  Let
$D(\delta t)$ be small at time intervals $\delta t$ larger or
comparable to the characteristic time $\tau_g$. Neither Gaussian
PDF nor $\delta$-correlation is assumed. The only important
assumption is that the processes $\rho_{ij}$ are stationary. Then
in the limit $t\gg 1$ the statistical moments of vorticity behave
as
$$
\langle |\omega|^n \rangle \propto e^{\Lambda_n t} .
$$ 
The limit of small values $n$ ($\Lambda_n \ll
\tau_g^{-1}$) corresponds to a $\delta$-correlated process. In
this limit $\Lambda_n$ obeys a linear differential equation of no
more than $n$-th order with constant coefficients.
We have calculated the values $\Lambda_n$ up to $n\sim 100$. 
They appeared to  satisfy approximately the 
relation
\begin{equation}
\label{4/3} \Lambda_n \simeq \alpha n^{4/3}
\end{equation}
(For exact values of first score of $\Lambda_n$ see \cite{PRL}.)
The nonlinear dependence of $\Lambda_n$ (\ref{4/3}) is the
manifestation of intermittency: $\langle \omega^{2n} \rangle \gg
\langle \omega^{n} \rangle^2$. In the limit of large $n$
($\Lambda_n \gg \tau_g^{-1}$) we find
$$  \Lambda_n \simeq \lambda n \ .$$
This means that at large values of $n$ intermittency breaks up. This
allows to find the structure functions of different orders
independently. The detailed discussion of the properties of the
equation (\ref{omega}) see in \cite{Wpechati}.

To illustrate the main ideas, consider a simple model of axially
symmetric, non-curved vortex filament with uniform vorticity
distribution inside its radius $R$. It contains all the main
features of a vortex filament. The difference from the general
case is  neglecting the curvature and precession of a vortex
filament and the gradient of vorticity inside it. We shall see
below that the first approximation corresponds to the late stages
of a vortex filament's evolution.

We seek a solution to the Euler equation in the linear form
\begin{equation}\label{tr}
v_{\phi} = \omega(t) r\,, \quad v_r = a(t) r\,,\quad v_z = b(t) z
\end{equation}
The  corresponding pressure is \cite{PRL}
$$
p(r,z,t) = P_1(t) r^2/2 + P_2(t) z^2/2
$$
From the Euler equation we then find differential equations for
$\omega$,$a$ and $b$. Combining them, we obtain \cite{PRL,JETP2}
$$ \ddot{\omega} = -P_2(t)\, \omega $$
This equation  corresponds to (\ref{omega}).
 In this model the growth of the 
 moments is easy
to understand: the amplitude of a solution remains roughly constant
for positive $P_2$ and grows exponentially if $P_2$ is negative.
Hence, on average the vorticity grows exponentially. This means
stretching of the vortex filament. From the Euler equation for
(\ref{tr}) it follows that $v_r/r=a=-\dot{\omega}/(2\omega)$, hence
the radius of a particle's orbit decreases as $r(t)\propto
\omega^{-1/2}$.

We note that the cross radius $R$ (\ref{biggradient}) of a vortex
filament restricts the region where vorticity is roughly constant.
The velocity $\omega R$ on the boundary of the region must be less
than the large-scale characteristic velocity pulsations $V_0$: on energetic
reasons, the vorticity stops growing in the regions where it
becomes comparable. Hence, the equation (\ref{omega}) is linear only
inside the radius $R\sim V_0/\omega$; to find the solution outside
$R$, feedback effect on $\rho_{ik}$ should be taken into account. As
the vorticity increases on average as a function of time, the radius
of a filament decreases, $R\sim 1/\omega$. Since the radii $r$ of
particles' orbits change as $1/\sqrt{\omega}$, we have $R/r\propto
1/\sqrt{\omega}$. 
This means that the
approximations of flat vorticity profile and of non-curved
axially-symmetric filament improve constantly as the filament
compresses. The vorticity continues to grow exponentially (on average)
until the cross  radius of the filament becomes comparable to the
viscous scale. Since we consider a stationary flow, vortex filaments
with large radii appear constantly, and at any time there is a set
of vortex filaments in different stages of their evolution.

Now we use this model to analyze the transversal structure functions
(\ref{SnEuler}).
 For simplicity, we assume circular orbits of particles in a filament.
Then
$$
{\bf v}({\bf r}) = {\bf r}\times {\bomega} \ , \qquad \delta \bv =
{\bf l}\times \bomega
$$
for any pair of points in the vortex filament separated by ${\bf
l}$. Averaging over all pairs of points in the  filament, we have $
\langle \delta \bv_{\perp}^n \rangle =l^n \omega^n$.

Now we must take the average over all vortex filaments. For given
$l$, the correlation breaks when $l$ becomes larger than the radius
$R$ of the filament:
\begin{equation}\label{break}
 l\geq R \sim V_0/\omega .
\end{equation}
Hence, calculating $S_n^{\perp}(l)$  we must account only the
filaments with $l\omega<V_0$.
This restriction for vorticity
produces the restriction for the filament's age $t_*$, which
results in the set of conditions:
$$  l^m \langle \omega^m \rangle  = l^m e^{\Lambda_m t_*} \le V_0^m
$$
for all $m$. Since $\Lambda_m/m$ grows as a function of $m$ and
$\Lambda_m /m \to \lambda$ as $m\to \infty$, we get
\begin{equation} \label{t*}
t_* = \inf\limits_m \left( \frac{m}{\Lambda_m} \ln \frac{V_0}{l}
\right) = \frac 1{\lambda} \ln \frac {V_0}{l} \ . \end{equation}
It
is natural that $t_*$ depends on high-order moments, since they are
more sensitive to rare events in intermittent media. Now the structure functions take the
form:
\begin{equation}\label{zetaE}
\begin{array}{c}
\displaystyle S_n^{\perp} (l) \simeq l^n \langle \omega^n \rangle (t_*) \propto
l^n e^{\Lambda_n t_*} \propto l^{\zeta_n} \ , \\
\displaystyle \zeta_n=n-\frac{\Lambda_n}{\lambda}
\end{array}
\end{equation}
One can see that the cutting parameter $V_0$ affects the amplitudes
of structure functions but not the scaling law. It means that the
boundary of nonlinearity affects on amplitude of structure functions
but not their scaling exponents. The vortex filaments with ages less
than $t_*$ have size larger then $l$ and contribute also to the
pre-exponents of structure functions.

This particular example reproduces the main aspects of the general
case. In the regions where vorticity is high the equation
(\ref{omega}) is linear, since $\rho_{ik}$  does not depend on
$\omega$. The boundary of these "linear" regions is restricted by
energetic reasons: in the pairs of points where velocity difference
becomes comparable to $V_0$ the equation (\ref{omega}) becomes
nonlinear, and the exponential growth of vorticity stops. So, the
relative velocity inside the vortex filament is restricted by the
condition
\begin{equation} \label{AV0}
\Delta v \le V_0 \ .
\end{equation}
The orthogonal component of velocity difference in two points
separated by $l$ inside a filament is $ \Delta {\bf v} = {\bf
l}\times {\bomega}$. \nopagebreak

 The condition (\ref{AV0}) then restricts the
age of the filament by (\ref{t*}), and averaging over all pairs of
points gives (\ref{zetaE}).

One can use analogous consideration to find the Lagrangian
structure function exponents. In two near points of a particle's
trajectory separated by time interval $\tau$ the velocities differ
roughly by $ \Delta v \simeq r_0 \omega^2 \tau $, where $r_0$ is
the parameter corresponding to momentarily curvature radius of the
trajectory. The linear equation (\ref{omega})  and exponential
growth of vorticity are only valid for small relative velocities.
The condition $\Delta v <V_0$ then gives $t\le t_*^L$, where
$$  \tau^m \langle \omega^{2m} \rangle  = \tau^m e^{\Lambda_{2m} t_*^L} \le V_0^m
$$
for all $m$.  Making use of 
$\lim\limits_{m\to\infty} (\Lambda_m/m)=\lambda<\infty$, we find
$$ e^{2\lambda t_*^L }=V_0/\tau \ ,$$
so the Lagrangian structure function is
\begin{equation}\label{zetaL}
\begin{array}{c}
\displaystyle S_n^L(\tau) = \langle \Delta v^n\rangle \propto \tau^n \langle
\omega^{2n}(t_*)\rangle \propto
\tau^{\xi_n} \ , \\
\displaystyle \xi_n=  n-\frac{\Lambda_{2n}}{2\lambda}
\end{array}
\end{equation}

In \cite{PRL} we obtained a similar expression for $S_n^L(\tau)$. It is
identical to (\ref{zetaL}) if we choose $\lambda=\Lambda_{4}/2 \simeq 3.1$.
The derivation in \cite{PRL} used the assumption that the structure function
$S_2$ was determined by stationary part of the probability density function.
In this paper we take the feedback effect into account. This allows to avoid the
assumptions of stationarity.

We do not consider longitudinal Eulerian structure functions in
this paper. To calculate them, one has to make a complicated
analysis of the stochastic equation (\ref{quasi}) and to
investigate the statistical behavior of $b_{ij}$, in addition to
$\omega$. From (\ref{quasi}) one can see that some linear
combinations of $b_{ij}$ (that do not  contribute to the
right-hand side of the second equation in  (\ref{quasi})) are
proportional to $\omega$. Then the longitudinal scaling exponents
may coincide with the transverse ones.

Basing on relations (\ref{zetaE}) and (\ref{zetaL}) one can find the
bridge relation between Eulerian transverse and Lagrangian scaling exponents:
\begin{equation}\label{bridge}
\Lambda_{2n}\left( n-\zeta_n \right)
=2\Lambda_n\left(n-\xi_n\right)
\end{equation}

Let us now compare  the predictions of our theory with  the results
of recent numerical simulations
\cite{Benzi} performed for $Re_{\lambda}\simeq 600$. This is the
only DNS that presents both Lagrangian and Eulerian transverse
scaling exponents. The scaling exponents are  normalized by
$\xi_2$ and $\zeta_3$, respectively (so-called ESS procedure). Our
theory has the only adjusting parameter $\lambda$. We use it to
fit the Lagrangian exponent $\xi_4/\xi_2$. Then we calculate all
other values $\zeta_n$ and $\xi_n$ from (\ref{zetaE}) and
(\ref{zetaL}). The results are presented in Table~1. One can see
that the theory is in excellent agreement with the experimental
data. \nopagebreak

Fig.1 presents the Eulerian scaling exponents (normalized by ESS)
as a function of the order $n$. The theoretical prediction is
presented together with experimental data taken from \cite{Benzi}
and \cite{Gotoh}. We see that, for high $n$, the longitudinal
scaling exponents correspond to the theory better than the
transversal ones. This is in accord with opinion of the authors of
\cite{Benzi}, who maintain that the results for high-order
transversal exponents are less certain.
\begin{center}
Table 1.  {\small Scaling exponents normalized by $\xi_2$ in
Lagrangian and by $\zeta_3$ in Eulerian case. The  results of
numerical simulations  are cited from \cite{Benzi}. }  
\begin{tabular}{|c|cc|cc|}  
  \hline
n  &   \multicolumn{2}{c|}{$\xi_n/\xi_2$ (Lagrange)} &
\multicolumn{2}{c|}{$\zeta_n/\zeta_3$ (Euler)}  \\
  &   DNS  &  Theory & DNS  & Theory \\
\hline
2&  &  &   $0.71\pm 0.01 $ & 0.72 \\
4& $1.66\pm 0.02$ & 1.66 &  $1.26\pm 0.01 $ &1.28 \\
6&  $2.10\pm 0.10$ & 2.14 &  $1.68\pm 0.03$ & 1.74\\
8&  $2.33\pm 0.17$ & 2.45 &  $ 1.98\pm 0.10$  &2.13 \\
10&  $2.45\pm 0.35$ & 2.64 & $2.25\pm 0.15$ & 2.46 \\
\hline
\end{tabular}
\end{center}

We stress that, to determine 
$\lambda$, we used the fourth-order Lagrangian scaling exponent.
So the theoretical prediction for Eulerian structure functions,
which is presented in Fig.1, has no adjusting
parameters. We also note that the concavity of the curve 
corresponds to intermittency of scaling exponents. In our theory
it is a direct consequence of intermittency of 
$\langle\omega^n\rangle$.
\vspace{-0.8cm}
\begin{figure}[h]
\includegraphics[width=7cm]{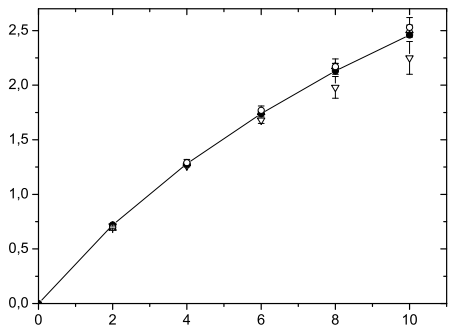}
\vspace{-0.8cm}
\caption{
 Eulerian relative scaling exponents $\zeta_n/\zeta_3$. Theory
($\bullet$ + line) and  experiment: transversal exponents taken
from \cite{Benzi} ($\bigtriangledown$) and longitudinal exponents
from \cite{Benzi} ($\bigtriangleup$) and \cite{Gotoh}
($\bigcirc$).   }
\end{figure}

The expressions (\ref{zetaE}) and  (\ref{zetaL}) allow not only
link the Lagrangian and Eulerian structure functions of the same
order, but also to link the scaling exponents of different orders.
Namely, using  (\ref{zetaE}) and  (\ref{zetaL}) we express
$\lambda$  as a function of $\xi_n/\xi_2$ and of $\zeta_n/\zeta_3$
independently for each n.
Substituting the scaling exponents obtained in \cite{Benzi}, we
find the values of $\lambda$ (see Table 2). We see that all the
values $\lambda$ obtained from both Lagrangian and Eulerian
scaling exponents for all measured $n$ coincide up to the
experimental errors.

To summarize, we report in this letter of the first substantial
advance in derivation of both Lagrangian and Eulerian scaling
exponents basing on the Navier-Stokes equation. We started from
the eq. (\ref{omega}), which is a direct consequence of the
Navier-Stokes equation in quasi-Lagrangian reference frame. We
treat it as a stochastic equation, assuming that $\rho_{ij}$ is a
stationary random process. This assumption is natural, since
$\rho_{ij}$ is independent of local vorticity.  Another important
assumption is that validity of the growing solutions of
(\ref{omega}) is restricted by energy limitation, which means that
velocity differences must be smaller than a large-scale eddy's
velocity. Analyzing solutions of (\ref{omega}), we find Lagrangian
and Eulerian transversal scaling exponents.
\begin{center}
Table 2.  {\small Free parameter $\lambda$ calculated 
from the relative scaling exponents $\zeta_n/\zeta_3$
(Euler) and $\xi_n/\xi_2$ (Lagrange) }\\ 
\begin{tabular}{|c|cc|}  
  \hline
n  &   \multicolumn{2}{c|}{$\lambda$} \\
   &    Lagrange &  Euler \\
\hline
4& \ $3.76 \pm 0.13$\  & \ $3.8 \pm 0.4$ \ \\
6& \ $3.67 \pm 0.23$ \ & \ $3.5 \pm 0.3$ \ \\
8& \ $3.57 \pm 0.26$ \ & \ $3.4 \pm 0.3$ \ \\
10& \  $3.54 \pm 0.35$ \ & \ $3.4 \pm 0.3$\  \\
\hline
\end{tabular}
\end{center}

We stress that the theory developed in this paper has  only one
fitting parameter. No additional propositions were used in
calculating both the scaling exponents. Neither dimensional nor
phenomenological hypotheses were used.  All the obtained
relations are  consequences of the equation (\ref{omega}) derived
directly from the Navier-Stokes equation. Thus, the coincidence
between the theoretical predictions, experimental results and
numerical calculations is very good.

We are very much obliged to Prof. A.V. Gurevich for his permanent
interest to our work and constant support. We are grateful to A.S.
Il'in for valuable remarks.

The work  was partially supported by the RAS Program "Fundamental
Problems of Nonlinear Dynamics".

\end{document}